\documentclass[10pt,conference]{IEEEtran}
\IEEEoverridecommandlockouts
\usepackage{cite}
\usepackage{amsmath,amssymb,amsfonts}
\usepackage{algorithmic}
\usepackage{graphicx}
\usepackage{textcomp}
\usepackage{xcolor}
\usepackage{caption}
\usepackage{tabu}
\usepackage{svg}
\usepackage{url}
\usepackage{tabularx}
\usepackage{blindtext}
\usepackage{subcaption}
\usepackage{longtable}
\usepackage{array} 
\usepackage{hyperref}
\usepackage{booktabs}

\usepackage{color, colortbl}
\usepackage{adjustbox}
\definecolor{Gray}{gray}{0.9}

\usepackage{tikz}

\title{Exploring Fairness Interventions in Open Source Projects
}
\author{
    \IEEEauthorblockN{Sadia Afrin Mim\IEEEauthorrefmark{1}, Fatema Tuz Zohra\IEEEauthorrefmark{2}
     Justin Smith\IEEEauthorrefmark{3}, Brittany Johnson \IEEEauthorrefmark{4}\\
    \IEEEauthorblockA{\IEEEauthorrefmark{1,}\IEEEauthorrefmark{2,}\IEEEauthorrefmark{4}Department of Computer Science, George Mason University}
    \IEEEauthorblockA{\IEEEauthorrefmark{2}Department of Computer Science, Lafayette College}
    \IEEEauthorblockA{
    \IEEEauthorrefmark{1}safrinmi@gmu.edu,
    \IEEEauthorrefmark{2}fzohra@gmu.edu,
    \IEEEauthorrefmark{3}smithjus@lafayette.edu,
    \IEEEauthorrefmark{4}johnsonb@gmu.edu
    }
    }
    }

\usepackage{bibentry}
\begin{document}

\maketitle

\begin{abstract}
The deployment of biased machine learning (ML) models has resulted in adverse effects in crucial sectors such as criminal justice, and healthcare. To address these challenges, a diverse range of machine learning fairness interventions have been developed, aiming to mitigate bias and promote the creation of more equitable models.
Despite the growing availability of these interventions, their adoption in real-world applications remains limited, with many practitioners unaware of their existence. To address this gap, we systematically identified and compiled a dataset of 62 open source fairness interventions and identified active ones. We conducted an in-depth analysis of their specifications and features to uncover considerations that may drive practitioner preference and to identify the software interventions actively maintained in the open source ecosystem. Our findings indicate that 32\% of these interventions have been actively maintained within the past year, and 50\% of them offer both bias detection and mitigation capabilities mostly during inprocessing.
\end{abstract}

\section{Introduction}

Machine learning (ML) applications and frameworks are having a major impact on modern society. In safety-critical domains such as healthcare, automotive, and manufacturing, ML systems have surpassed human performance by processing large amounts of data and making predictions or classifications with greater reliability~\cite{iqbal2022assurance}. 
Although these advances are promising, numerous studies have also revealed that many ML applications are prone to undesirable bias. Examples include, but are not limited to, biases in the evaluation of physical appearance~\cite{mehrabi2021survey}, demographic biases in loan application approvals~\cite{lipsitz2006possessive}, gender bias in employee recruitment~\cite {fitria2021gender}, natural language processing~\cite{hasan2024olf} and racial bias in determining criminal sentencing~\cite{fitria2021gender}.

In response to growing concerns regarding bias in machine learning models, researchers and practitioners have proposed a variety of potential solutions.
The most common form of support provided is automated software interventions designed to help identify, mitigate and evaluate bias within machine learning models and their training datasets~\cite{brun2018software, lee2021landscape, johnson2021towards}. 

The availability of fairness interventions provides practitioners with an option for addressing fairness concerns in practice, however, research indicates that practitioners may lack the necessary expertise to use them effectively~\cite{lee2021landscape}. 
Furthermore, as the range of interventions continues to expand, the challenge of identifying and evaluating the most suitable options has also become complex.

\textit{The goal of our research is to provide a comprehensive exploration of open source fairness interventions to form foundations that can simplify thieir selection and evaluation process.} 
To this end, we conducted an in-depth analysis of currently available interventions, focusing on their features and specifications for use. 
We compiled a collection of available, fully functional fairness interventions and analyzed the information provided in their documentation and tutorials.
Additionally, we determined which interventions are \textit{active} and  \textit{inactive} based on their recent commit activity. 
We further explored the underlying ecosystems, algorithmic fairness approaches, domain-specific applications, and mitigation techniques for each intervention to better understand the ways in which these interventions might meet the needs of practitioners. The contributions of this work are:
\begin{itemize}
    \item An extensive compilation of 62 open source fairness interventions available for practitioners to incorporate into their workflows.
    \item A qualitative assessment and categorization of the features and specifications offered by existing interventions that can support their adoption and engagement in the open source community.
\end{itemize}

The remainder of this paper is organized as follows: Section \ref{related} reviews the related literature. Section \ref{method} outlines our methodology, beginning with the formulation of research questions that guided our analysis of fairness interventions, including their features, specifications, documentation, and tutorials. Section \ref{finder} presents the results of our analysis. Insights derived from these findings are discussed in Section V and some possible threats to the validity of our work are in Section \ref{validity}. We conclude the paper in Section \ref{conclusion}.

\section{Related Work}
\label{related}

Numerous efforts have focused on exploring various aspects of software interventions and machine learning fairness. Relevant to our work, prior research has examined the taxonomy of software interventions, provided insights into machine learning fairness, and proposed approaches for mitigating bias in machine learning models.

\subsection{Software Interventions Taxonomies}

Creating software intervention taxonomies is a common approach in the software engineering research community. Researchers have introduced various taxonomies over time to assist developers in selecting the interventions best suited to their tasks. Delgado and colleagues developed a structured taxonomy for runtime software-fault monitoring interventions ~\cite{delgado2004taxonomy}. They categorized the interventions based on key components such as specification languages, monitoring mechanisms, and event handlers. This taxonomy helps developers in understanding, classifying, and distinguishing advancements in monitoring approaches. Shaukat and colleagues developed a taxonomy to classify and compare automated software testing interventions ~\cite{shaukat2015taxonomy}. This taxonomy assists professionals in selecting the most appropriate interventions by analyzing 32 automated testing interventions based on attributes like operating system compatibility, programming language support, licensing, browser support, and cost. The work by Lee and colleagues represents a significant contribution to the field of machine fairness intervention analysis ~\cite{lee2021landscape}. Although the paper does not explicitly offer a taxonomy, it provides a comparative analysis of six prominent open source fairness toolkits, such as IBM Fairness 360 (AIF360) and Google What-If intervention. The authors assessed the interventions by examining their features and limitations, including aspects such as fairness metrics, models, and bias mitigation algorithms. This analysis emphasizes functionality, usability, and applicability, highlighting the differences in the design, features, and target audiences of these toolkits.

Pagano and colleagues reviewed current research on bias and unfairness in machine learning (ML) models. They examined 45 studies published between 2017 and 2022 and organized information into categories such as types of bias (data, algorithm, user interaction), fairness metrics, datasets, and mitigation techniques (preprocessing, inprocessing, postprocessing) ~\cite{pagano2023bias}. It helps researchers select a intervention by providing detailed evaluations of interventions like AIF360, FairLearn, and Aequitas, outlining their features, limitations, and suitability for different bias identification and mitigation scenarios.

\subsection{Preliminary Concepts of Machine Learning Fairness}

Fairness in machine learning has gained significant importance as an essential topic to address as it important to develop systems that are unbiased and promote equity. Mehrabi and colleagues explored bias and fairness in machine learning, identifying key sources of bias and discussing their real-world implications~\cite{mehrabi2021survey}. They also reviewed mitigation strategies and fairness toolkits, providing valuable insights for addressing unfairness in ML systems. Zhang and colleagues investigated structural biases and limitations in fairness-aware algorithms, emphasizing the need for interventions that address both technical and sociopolitical biases in ML workflows ~\cite{zhang2021ignorance}. Makhlouf and colleagues examined how fairness in machine learning models interacts with the broader sociotechnical context, highlighting the impact of abstractions on fairness outcomes and pointing the importance of a contextual understanding of fairness beyond technical solutions ~\cite{makhlouf2020survey}. Caton and colleagues analyzed datasets, fairness metrics, and bias mitigation techniques, discussing open source libraries and how feature alteration and training data size influence fairness ~\cite{caton2020fairness}. Lee and colleagues categorized fairness-aware datasets by domain, such as healthcare, finance, and education, identifying challenges like dataset bias and imbalance while offering insights into their suitability for fairness research ~\cite{le2022survey}. Pessach and colleagues reviewed key sources of algorithmic bias, including biases in datasets, objectives, and proxy attributes, uncovering the roots of unfairness in ML systems ~\cite{pessach2022review}.

\subsection{Evaluations of Bias Detection \& Mitigation Techniques}

Addressing bias in machine learning is also essential to ensure equitable and ethical decision-making across various applications. Bias mitigation in machine learning has been explored through various empirical and methodological approaches. Biswas and colleagues conducted an empirical evaluation of approximately 40 machine-learning models from Kaggle, assessing five different tasks, including fairness metrics, to bridge the gap between theoretical fairness concepts and practical applications in software engineering ~\cite{biswas2020machine}. Hort and colleagues proposed the Fairera model, demonstrating that bias mitigation algorithms were ineffective in a significant number of cases and highlighting the trade-offs between fairness and accuracy that often arise after applying these algorithms ~\cite{hort2021fairea}. Researchers have also focused on specific stages of bias mitigation: Wan and colleagues reviewed inprocess bias mitigation techniques ~\cite{wan2023processing}, while Lohia and colleagues concentrated on postprocessing approaches ~\cite{lohia2019bias}. Domain-specific efforts include bias mitigation for face recognition systems ~\cite{wang2020mitigating} and the medical sector ~\cite{noseworthy2020assessing}. Additionally, Amini and colleagues introduced an approach to derive bias mitigation algorithms by leveraging the structure of datasets ~\cite{amini2019uncovering}.

Based on previous research exploring various facets of fairness interventions and techniques in open source machine learning, prior work has primarily focused on specific interventions, taxonomies, and bias mitigation approaches. However, much of this research tends to target individual interventions or smaller sets of interventions or stages of bias mitigation rather than providing a comprehensive analysis of fairness interventions in the broader open source ecosystem. The most closely related work to our own is that of Lee et.al.~\cite{lee2021landscape}. While the study by Lee and colleagues focuses on six prominent open source fairness toolkits, our work provides a more comprehensive analysis, covering a significantly larger dataset of 62 interventions. Unlike prior work, our analysis delves into various critical aspects, such as categorizing interventions by domain type, determining their active or inactive status, conducting phase-specific bias mitigation evaluations, and examining platforms and functionalities. Furthermore, our study distinguishes between interventions that function as analysis interventions versus those offering guidance, presenting a broader and more detailed understanding of the fairness intervention landscape.

\section{Methodology}
\label{method}

Our research aims to better understand and characterize the landscape of machine learning fairness interventions available for practitioners. 
We gathered and examined intervention name-specific data to answer the following research questions:

\begin{description}
   
    \item[\textbf{RQ1}] \textit{In what ways are open source fairness interventions made available for integration and use?}
    \item[\textbf{RQ2}] \textit{What machine learning algorithms do existing open source fairness interventions work with or support?}
   
    \item[\textbf{RQ3}] \textit{What are factors that can help distinguish existing fairness interventions in terms of the kind of support they provide?}

    \item[\textbf{RQ4}] \textit{To what extent do these interventions support bias detection and mitigation across the different stages of the machine learning lifecycle?}
\end{description}

\subsection{Fairness Intervention Dataset}
To answer our research questions, we started with a sample of 10 fairness interventions curated in prior work~\cite{mim2023taxonomy}. 
We expanded this set using keyword-based mining to identify additional interventions on GitHub that appeared to offer fairness-related support for machine learning. 
We analyzed the information provided in the \textbf{About} section of each intervention repository to identify commonly used terms that could be used  as keywords for our search.

Our final set of keywords included keywords/phrases such as ethics, ML fairness, ML ethics, ethical ML, machine learning bias, AI bias, AI fairness, fairness metrics, bias mitigation, fairness algorithm, and fairness toolkit.

We developed Python scripts using the GitHub API\footnote{\url{https://docs.github.com/en/rest?apiVersion=2022-11-28}} to perform our keyword-based searches. 
We manually analyzed the returned repositories, with the most relevant interventions generally found within the top 5~6 results. To ensure comprehensive coverage, we collected the top 10 results for each keyword. In total, we curated a list of 62 interventions, which is publicly available online~\footnote{\url{https://github.com/INSPIRED-GMU/Qualitative-Analysis-of-Fairness-Tools}}.

\subsection{Determining Active Status}
To determine the active status of each intervention in our dataset, we used a model from prior work that classifies open source repositories into two categories based on repository activity~\cite{coelho2020github}.
We classified repositories as \textit{active} based on repository commit activity within the past year.
We classified repositories as \textit{inactive} if the repository had no commit activity in over a year. We also considered repositories marked as read-only or archived by their owners as \textit{inactive}~\cite{de2019identifying}.

\subsection{Evaluating Availability \& Compatibility (RQ1)}

To answer \textbf{RQ1}, we analyzed repository documentation, including the \texttt{README} files and information in the \texttt{About} section. 
We also collected and analyzed any available research literature related to each interventions.
We considered the following aspects of intervention availability and compatibility:

\begin{enumerate}
    \item \textbf{Language Compatibility.} For this, we looked for the language(s) each intervention works with. We typically found this information in the source code and the `Languages' information in the `About' panel for each repository (e.g., \textsc{Aequitas})~\footnote{\url{https://github.com/dssg/aequitas}}.
    \item \textbf{Platform Availability.} 
    To determine platform availability, or how the support is made available for use, we analyzed the documentation, README files, and `About' section for any information on the way the features and functionality are provided. 
    For example, in \textsc{LiFT} repository~\footnote{\url{https://github.com/linkedin/LiFT}} `About' section, the intervention is labeled as a library but can also be used as a toolkit as indicated by the name (LinkedIn Fairness toolkit).
    We found the following ways in which fairness interventions make themselves available for integration into practitioner workflows:
\begin{enumerate}
    \item \textbf{Tool.} A stand-alone program or intervention created for a particular purpose or to support the completion of a specific task.
    \item \textbf{Toolkit.} A group of tools that can be used to tackle a more diverse set of general tasks~\cite{jaffe2024toolkit}.
    \item \textbf{Package.} A collection of modules and code that are bundled for reuse and distribution where the user is in charge of the flow~\cite{abadi2016tensorflow}.
    \item \textbf{Library.} A collection of pre-written, reusable packages that can be used to support the completion of certain kinds of programmatic tasks.~\cite{jones2000composing}.
    \item \textbf{Framework.} A structured foundation that provides architecture and workflows for building applications where the framework itself is in charge of the flow~\cite{cwalina2020framework}.
\end{enumerate}

    \item \textbf{License Availability.} This information refers to if and how the interventions are licensed. 
    We found this information in the `About' portion where the license is listed (e.g., \textsc{fairlearn}~\footnote{\url{https://github.com/fairlearn/fairlearn}})
\end{enumerate}

\subsection{Evaluating Algorithm Coverage (RQ2)}

To answer \textbf{RQ2}, we analyzed repository and intervention information to determine the range of algorithms each intervention provides support for.
To accomplish this, we conducted a comprehensive analysis of README files to understand the functionality and intended use cases, source code to identify implemented fairness techniques and algorithms, and related research literature for additional insights into algorithmic support.
This mixed methods approach allowed us to triangulate information and develop a more complete understanding of the algorithms supported. 
For example, only by looking at \textsc{fairkit-learn}'s related research paper~\cite{johnson2022fairkit} did we learn that the intervention works with regression, adversarial, and random forest algorithms.

\subsection{Analysis of Distinguishing Factors (RQ3)}

To answer \textbf{RQ3}, we analyzed the information we curated for each intervention to determine themes, other than the kinds of fairness support provided, that might help practitioners distinguish and decide on appropriate (and useful) fairness interventions.
Based on an analysis of repository artifacts and related literature (if available), we determined three different possible considerations: (1) whether the intervention supports bias detection or mitigation (or both), (2) whether the intervention provides guidance or analytic support, and (3) whether the intervention provides generic or special purpose support. 

An intervention that supports \textbf{bias detection} allows practitioners to determine if and to what extent various kinds of bias exist within a given model or dataset, most often using fairness metrics.
One that supports \textbf{bias mitigation} leverages algorithms and techniques to reduce the likelihood of bias when using a given model and/or dataset. 

Interventions that provide \textbf{guidance} support provide some general insights into fairness considerations for a dataset or model without conducting any analyses or calculating any fairness scores. 
One example of this is \textsc{Deon}, which provides an ethical checklist to ``provoke discussion'' that can lead to action.~\footnote{\url{https://github.com/drivendataorg/deon}}
Interventions that provide \textbf{analytic} support perform automated analyses that detect or mitigate dataset or model bias. The includes the use of fairness metrics to evaluate model fairness. 
An example of an analytic intervention would be \textsc{langtest}, where the repository artifacts and information suggest it can be used to test for bias and fairness-related concerns.~\footnote{\url{https://github.com/JohnSnowLabs/langtest}}

Lastly, a \textbf{generic} intervention provides support that is domain agnostic. For example, \textsc{AI Fairness 360} does not mention any specific domain for which the intervention is designed to support (or not support) and would therefore be considered a generic intervention.
\textbf{Special purpose} interventions provided some indication that the intervention was meant to be used in niche domains (e.g., healthcare) or for specific technologies (e.g., large language models).
For example, the repository for \textsc{fairMLHealth} states that the intervention is ``healthcare-specific''.~\footnote{\url{https://github.com/KenSciResearch/fairMLHealth}}

 \subsection{Analysis of ML Lifecycle Support (RQ4)}

To answer \textbf{RQ4}, we analyzed repositories to determine mappings between the support provided by the intervention and the various phases involved in the machine learning process. 
To organize and simplify how we will report our findings, as with prior work~\cite{mim2023taxonomy}, we group these steps into three phases: \textbf{\textit{preprocesing}}, \textbf{\textit{inprocessing}} and \textbf{\textit{postprocessing}}. Preprocessing reduces bias by modifying data before training, inprocessing integrates fairness constraints during model training, and postprocessing adjusts outputs after training to promote fairness across groups\cite{caton2024fairness}.
To determine which phase(s) each intervention provides support for, we analyzed intervention documentation, README files, and related research for any relevant insights. 
For example, in the \textsc{AI Fairness 360} repository, the  documentation mentioned optimized preprocesing, prejudice remover inprocessing , and equalized odd postprocessing algorithms.

\subsection{Analysis \& Result Verification}

For each research question, we wanted to ensure that the data collected and analyzed was both relevant and accurately interpreted for our categorizations. 
Therefore, using the labels and categories outlined above for each research question, two authors independently analyzed each intervention in our dataset. 
In this phase, each author drove their own analysis based on their own list of relevant information and outcomes for each research question.
Once each author finished their analysis, we then collaboratively discussed and evaluated these results, identifying and addressing any inconsistencies or disagreements. To address these discrepancies, we revisited the intervention and categorization  where disagreements occurred and relabeled them collaboratively. 
At the end of this iterative process, we had a final set of labels for each intervention with respect to each research question which we report next.

\section{Findings}
\label{finder}

\begin{figure*}[ht]\centering
  \includegraphics[width=12cm,height=6cm]{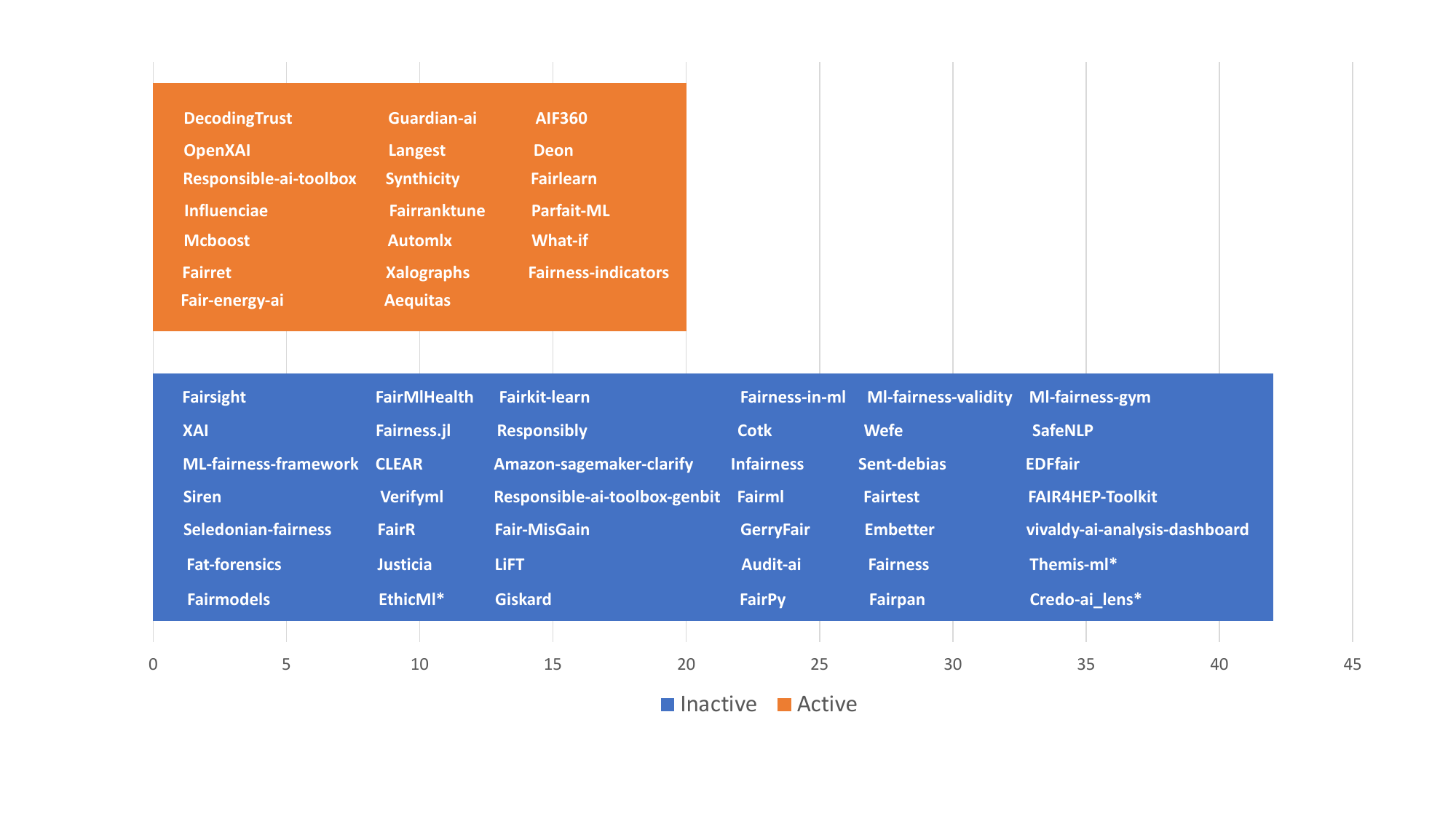}
  \caption{Available Fairness interventions in Open Source}
  \label{active}
\end{figure*}

Our efforts have uncovered valuable insights regarding the landscape of existing open source fairness interventions. 
This section presents our findings, namely intervention availability and compatibility (\textit{RQ1}), coverage of algorithms (\textit{RQ2}), distinguishing factors (\textit{RQ3}), and machine learning lifecycle support (\textit{RQ4}).
Among the 62 open source fairness interventions in our dataset, we found that only 32\% are being actively maintained as of 2024. In Figure~\ref{active}, we list the projects that are considered active and inactive, respectively. We denote archived repositories with an asterisk (*).

\subsection{Fairness intervention Availability \& Compatibility (RQ1)}
\begin{figure}[ht]
  \includegraphics[width=5cm]{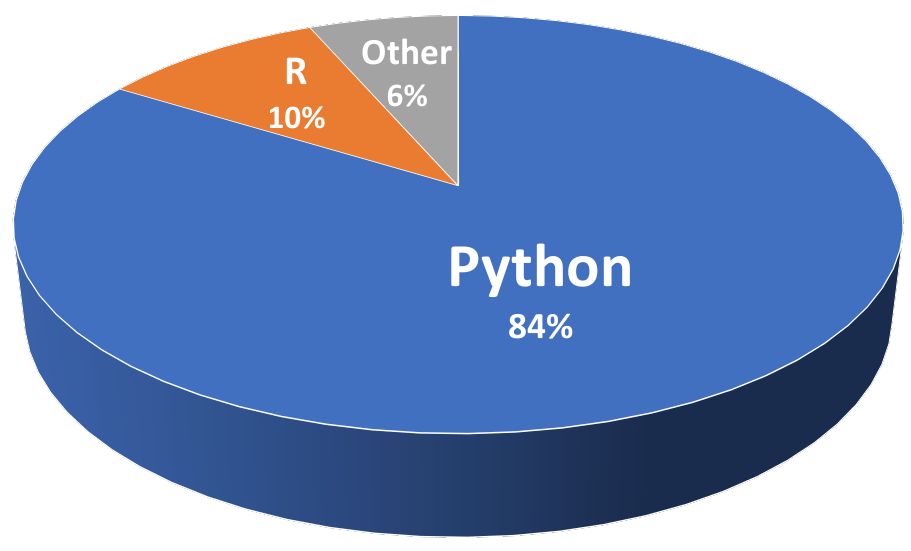}\centering
  \caption{Programming Language Supported Across Interventions} 
  \label{language}
\end{figure}

With respect to \textit{language availability}, we found that the interventions in our dataset utilized or supported the following languages: Python, R, Julia, Scala/Spark, and Svelte.
While we found a range of languages across intervention, in the set of active interventions Python and R were the only languages supported (with Python being the most prevalent as shown in Figure~\ref{language}). 

\begin{figure}[ht]
  \includegraphics[width=7cm]{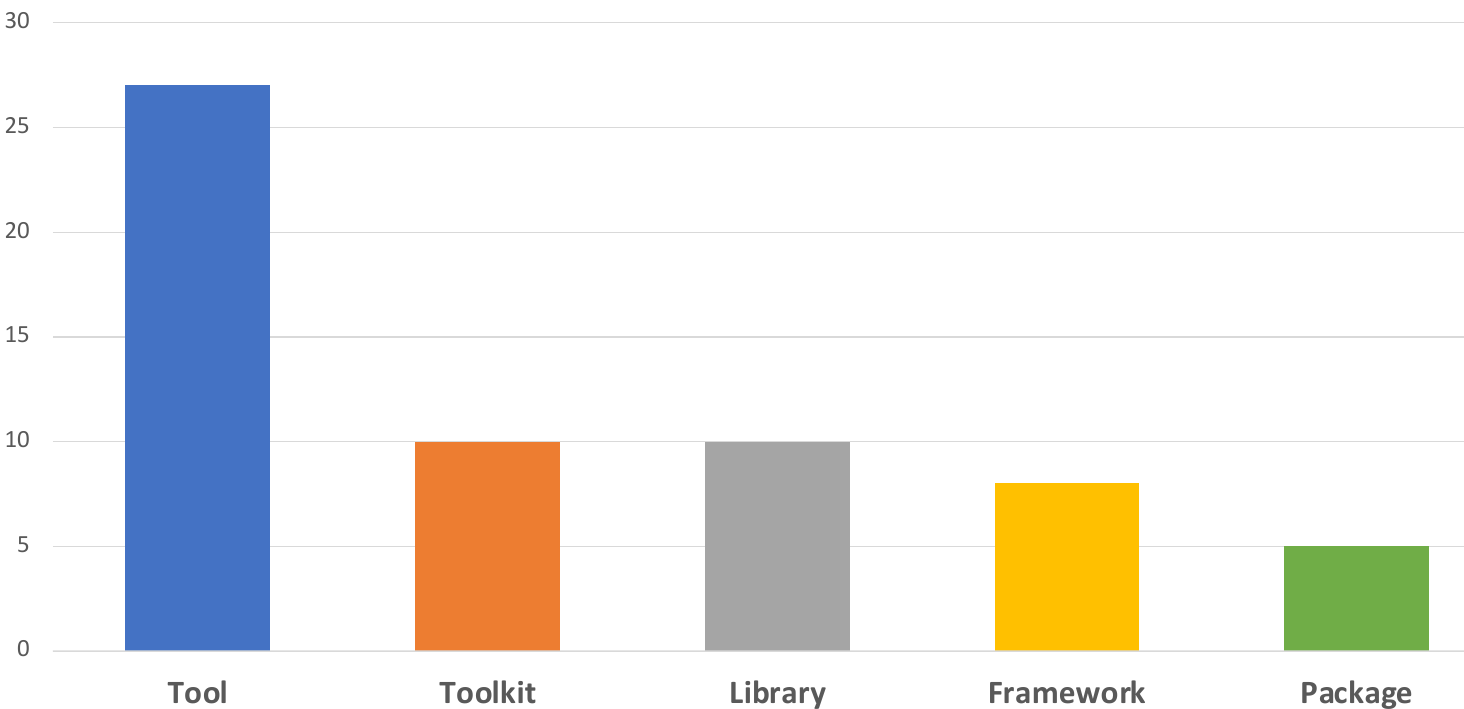}\centering
  \caption{Distribution of different Fairness Software Interventions} 
  \label{platform}
\end{figure}

For \textit{platform availability}, we found several ways  in which interventions make themselves available for integration into practitioner workflows. We mentioned our software platform categories in the methodology section which are \textbf{Tool}, \textbf{Toolkit}, \textbf{Package}, \textbf{Library}, \textbf{Framework}.
As shown in Figure~\ref{platform}, active interventions are available for integration and use in a diversity of ways (though many presented themselves as interventions). 
We also found that a single intervention could sometimes be used (or described) in multiple, overlapping ways. 
For example, we found that often interventions would label themselves as toolkits but also note that the repository contained a ``library'' of support.

\subsubsection{Fairness Intervention Licensing}
\begin{figure}[ht]
  \includegraphics[width=8cm]{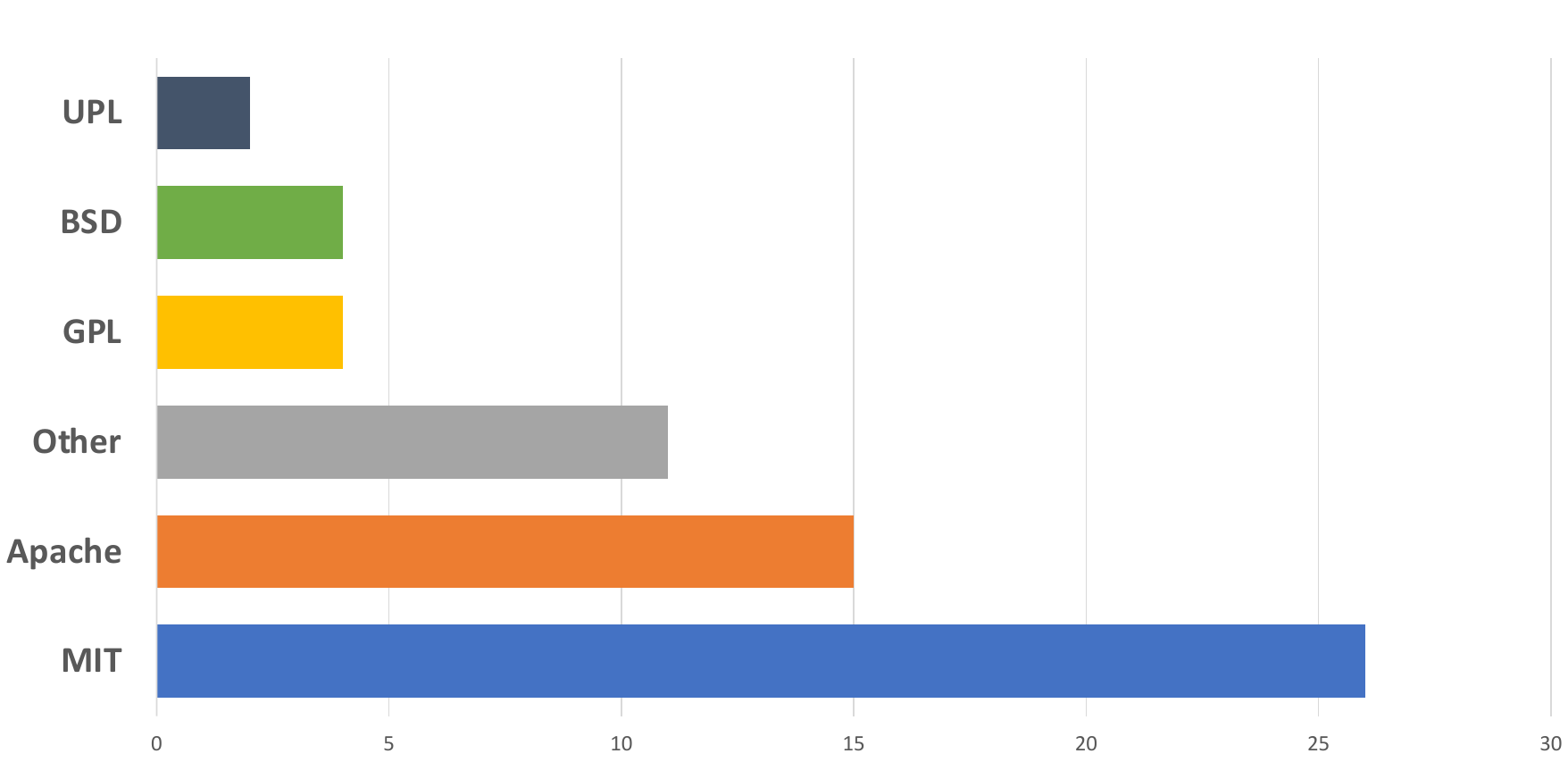}\centering
  \caption{Distribution of License Types} 
  \label{license}
\end{figure}

With respect to \textit{license availability}, as shown in Figure~\ref{license}, we mostly encountered the following licenses: MIT, Apache, Universal Permissive License (UPL), GNU General Public License (GPL) and Berkley Source Distribution (BSD), Others(CC-BY-SA-4.0, GNU, Barco Vivadly, AGPL, Custom License). 
The Other category includes those sets of licenses that have appeared only once or twice in the set of interventions.

According to our analysis, active interventions are most often licensed under either an MIT license or an Apache license. Which may be due to its permissive nature, allowing free use, modification, and distribution, fostering widespread adoption. 
In contrast, restrictive licenses like GPL or AGPL require derivative works to remain open source, limiting commercial use and adoption.

\subsection{Machine Learning Algorithm Coverage(RQ2)}
Our analysis found that existing interventions provide support for a wide range of machine learning algorithms.
The most common algorithms supported across interventions in our dataset include random forest classifier, generative pre-trained transformer (GPT), decision tree classifier, support vector machine, logistic regression, linear regression, binary classification, multiclass classification, reinforcement learning, least squares regression, and K nearest neighbor.
In some cases, interventions described the algorithms supported by specifying the class of algorithms it works with. Building on this insight, we found that the mostly commonly supported \textit{classes} of algorithms include classification, regression, natural language processing (NLP), large language models (LLMs), adversarial networks, and Bayesian networks. 
Our analysis indicates that a significant proportion of active fairness interventions are designed to address biases in classification models. This focus likely stems from the widespread use of classification algorithms in critical decision-making areas~\cite{lee2021landscape}. In our list of interventions, 52 of them work on ensuring fairness for classification algorithms, and 34 of them work on regressions. We identified 10 interventions that provide support for LLMs and NLP models.

\subsection{Distinguishing Factors (RQ3)}\label{subsec:rq3}

We examined the potential distinguishing factors between fairness interventions by focusing on three key aspects: whether the interventions are designed provided bias detection and/or mitigation support, whether interventions provided guidance or analytic support, and whether the interventions are designed for special or specific purposes.

\subsubsection{Detection \& Mitigation Support}

Out of the 62 interventions analyzed, 30 are focused on bias detection, meaning they only indicate and measure the presence of bias, while 32 also include bias mitigation capabilities. 

\begin{figure}[ht]
  \includegraphics[width=3.5cm]{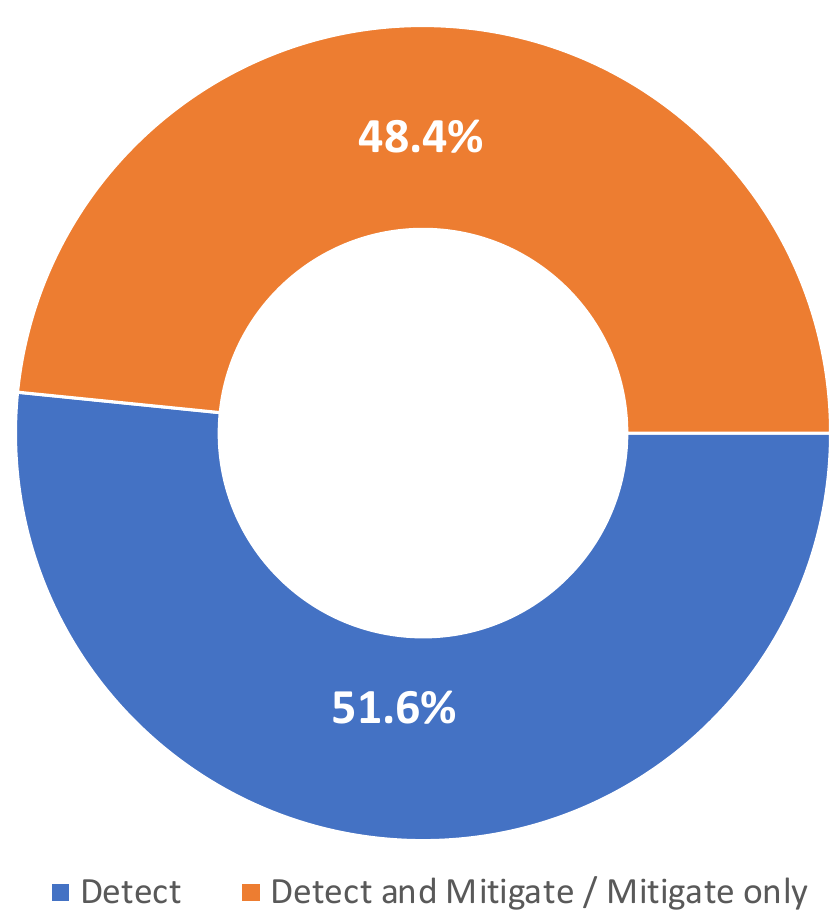}\centering
  \caption{intervention Distribution: Detect vs Detect and Mitigate} 
  \label{detect}
\end{figure}

For bias mitigation, the majority of these interventions implement mitigation algorithms at various stages of the machine learning lifecycle. In Figure~\ref{detect} we showed the availability of these categories in open source and our analysis reveals that half of the active fairness interventions are equipped to not only detect bias but also implement mitigation strategies. For bias detection, these interventions use various metrics to evaluate fairness, with some exhaustive ones e.g., AIF360, offering an extensive range of metrics—AIF360 alone includes 70 fairness measures. These metrics provide comprehensive insights into fairness, addressing both individual and demographic-level disparities

Individual fairness metrics include the Theil Index, Feature Agreement, Rank Agreement, and Graph Laplacian Individual Fairness. Group fairness metrics commonly observed are Equalized Odds, ROC Pivot(), Disparate Impact, Statistical Parity, Demographic Parity, Equal Opportunity, Balanced Group Threshold, and Exposure Rank. Among mitigation interventions, while some focus exclusively on bias mitigation, the majority integrate both bias detection and mitigation. For instance, the \textit{fairness-in-ml} intervention is dedicated solely to bias mitigation without incorporating bias detection or measurement using adversarial networks.

\subsubsection{Guidance \& Analytic Support}
Based on the functionality of these interventions, we categorized them into two groups: guidance and analysis. Our findings indicate that analysis interventions are more prevalent, with only 25\% of active fairness interventions offering guidance without explicit quantitative analysis. For inactive interventions, this proportion is even smaller, with just 11\% providing guidance. Since guidance interventions operate on more abstract concepts rather than empirical data, analysis interventions are better aligned with real-world demands. Consequently, the development of guidance interventions is also less frequent compared to analysis interventions.

 \subsubsection{Generic \& Special Purpose Support}
\begin{figure}[ht]
  \includegraphics[width=7.5cm]{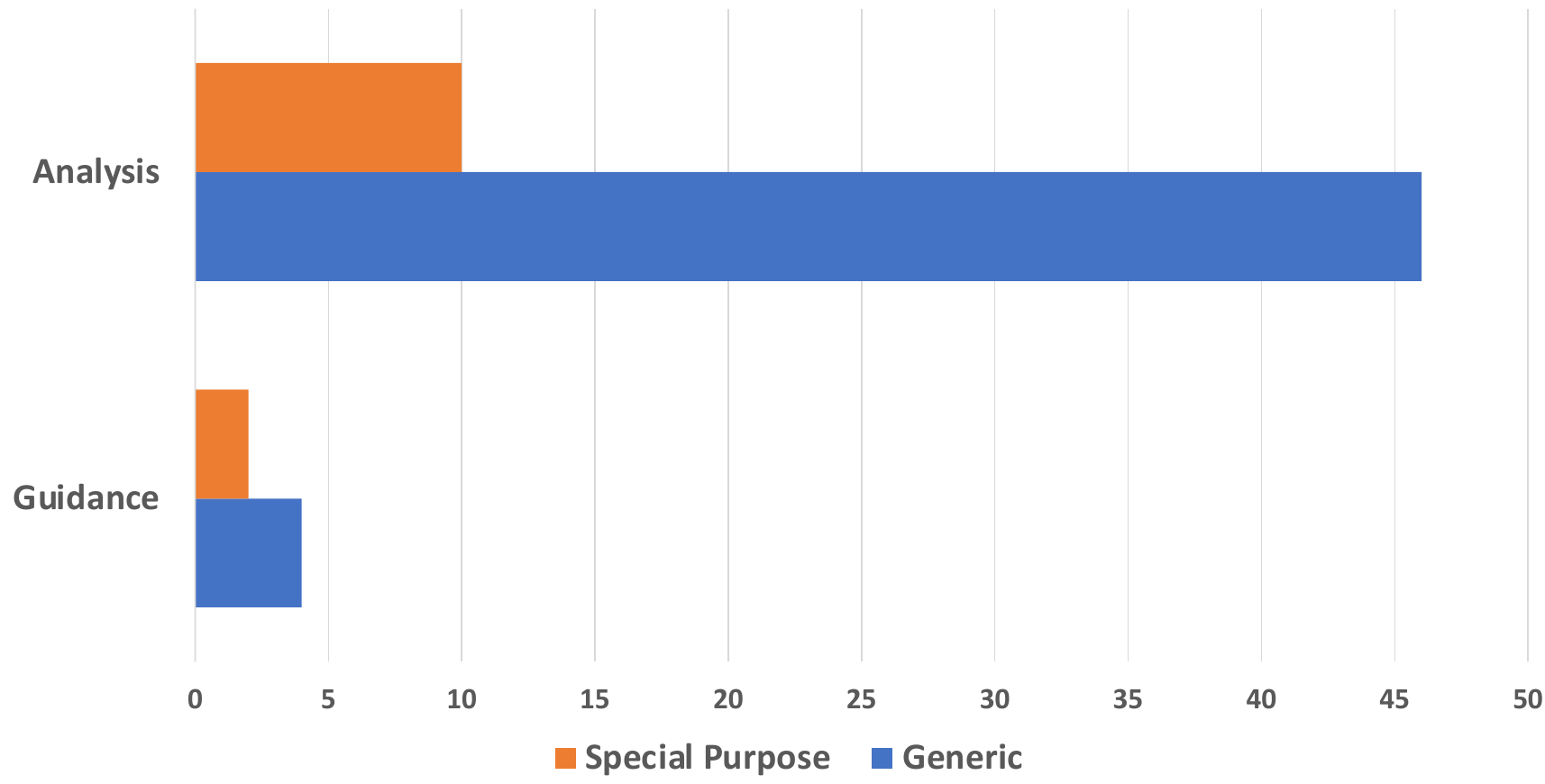}\centering
  \caption{Distribution of Generic and Special Purpose interventions for Guidance and Analysis Category} 
  \label{guidance}
\end{figure}

 As shown in Figure~\ref{guidance}, we found both generic and special purpose analytic and guidance interventions in terms of real-world application domain.
 The special purposes available for these interventions are healthcare, energy management, online news recommendation, GPT models, and word embedding. Our analysis revealed that generic interventions are highly prominent in our dataset. Among analysis interventions, 82\% are generic, as they do not specify any particular application area in their documentation. Similarly, even within the smaller subset of guidance interventions, 66\% are found to be generic. 
 
\subsection{Machine Learning Lifecycle Support (RQ4)} \label{subsec:rq4}

To address RQ4, we examined the support interventions in our dataset provide across the different stages of the machine learning lifecycle. We analyzed the types of fairness metrics and mitigation algorithms offered across various phases of the machine learning fairness process. However, the related literature and documentation lacked any classification of these metrics within the context of the machine learning life cycle. Therefore, we focused our analysis on  bias mitigation algorithms implemented during the preprocesing, inprocessing, and postprocessing phases, evaluating the type of support these algorithms provide for mitigating bias at each stage.

\subsubsection{preprocesing Support}

The identified technologies highlight the diverse strategies employed in the field of AI bias mitigation. \textit{Feature selection}, \textit{Reweighting}, \textit{Data Augmentation},\textit{Sampling},\textit{Encoder}, \textit{Imputer}, \textit{Seledonian Regression Technique},\textit{Disparate Impact Remover},\textit{ExponentiatedGradient}, \textit{AdversarialFairnessClassifier}, \textit{Relabeling}, \textit{Optimized Preprocessing}, \textit{Synthetic Data Generation}, \textit{Sensitive Set Invariance}, \textit{Balanced Data Splitting} are common preprocessing techniques. Advanced interventions combine multiple approaches to ensure comprehensive bias mitigation.

\subsubsection{inprocessing Support}  

The identified inprocess mitigation techniques can be categorized into adversarial approaches, additive models, and equalized odds with linear programming. Adversarial debiasing and training methods are prominently used to mitigate bias during the learning process by actively adjusting the model to counteract unfairness. The common approaches used by the interventions during inprocessing is \textit{Additive
Counterfactually Fair(ACF)}, \textit{Adverserial Debiasing}, \textit{Prejudice Remover Regulizer}, \textit{Adverserial Training}, \textit{Calibrated Equalized Odd}, \textit{Explicitly Deweighted Features},
\textit{Exponentiated Gradient Reduction}, \textit{Fair Rank Algorithm}, \textit{Meta Fair Classifier}.

\subsubsection{postprocessing Support}

The postprocess mitigation techniques identified include  \textit{Ranking Adjustments}, \textit{Calibration}, \textit{Adversarial Training}, \textit{Threshold Optimization}, \textit{Downstream operations}, \textit{Equalized Odds Postprocessing}, \textit{Reject Option Classification}, \textit{Adjusting Model Predictions}, \textit{Rich Subgroup Fairness}, \textit{Rich Subgroup Fairness}. Among Active interventions, inprocess mitigation techniques are the most commonly used approach, slightly ahead of preprocessing.
postprocess mitigation techniques are popular among the interventions that are classified as Inactive. 

 Active interventions provide more inprocessing support. Around 50\% of the active interventions include inprocess bias mitigation support. The Inactive interventions provide more preprocessing and postprocessing bias mitigation support, which is 40\% of total Inactive interventions. Another notable observation is that some algorithms are effective for bias mitigation across multiple phases. For instance, \textit{Adversarial Training} functions as both an inprocess and postprocess bias mitigation algorithm.

\section{Discussion}
\label{disussion}
Our findings help bring structure to the spectrum of available open source fairness interventions. Our efforts provide useful insights that can be leveraged to inform efforts aimed at providing model fairness support for practitioners.  We discuss these insights in detail next.  

\subsection{Supporting Practitioners Seeking Fairness intervention Support}

There is an ever-growing list of available intervention support for machine learning model fairness. While some practitioners may have existing knowledge of intervention support that they can leverage in their decision-making, research suggests that most practitioners may not even be aware of the various interventions available to them~\cite{lee2021landscape}. Furthermore, one of the challenges to adopting fairness interventions in practice is difficulty searching for and deciding among available interventions. The increasing number of interventions, along with the diversity of features and support provided (as  highlighted in our efforts, can lead to increased difficulty for practitioners when trying to find and integrate the most effective (and relevant) intervention for their environment and goals.

Our analyses found that most of the interventions in our dataset provide analytic support (Section~\ref{subsec:rq3}), but as emphasized in prior work (which most often focuses on evaluating analytic support) practitioners struggle with finding and making decisions about adequate analytic support~\cite{lee2021landscape}. 
While only few in number in comparison to analytic interventions, we found that there are interventions designed to provide guidance on making fairness considerations. 
This illuminates opportunities for future research and development on the potential value of guidance interventions and the kinds of guidance we can operationalize as interventions to support fairness considerations. For instance, expanding on the concept of interventions that display nearby samples to guide fairness to a particular biased scenario\cite{mcqueary2024py}.

\subsection{Improving Diversity in Available Fairness Intervention Support}

Our findings also further emphasize the large landscape of algorithms and metrics (Section~\ref{subsec:rq4}), which as prior works suggests can be overwhelming to reason about. 
However, we also provide a more comprehensive characterization of the landscape of existing machine learning fairness interventions that can provide foundations for improvement. 
More specifically, we describe ways to organize these kinds of considerations (e.g., by ML lifecycle phase) as well as other ways to help distinguish between existing interventions (e.g., providing support for a specific domain or class of algorithms). 
Perhaps a better understanding these kinds of distinguishing factors that may exist and that practitioners care about could help improve the ability navigate the space of interventions and better facilitate use in practice.

In building this foundation, we also observed potential gaps in existing support. 
For example, one of the areas where we found the least support in our categorization was in handling diverse and complex data types such as those necessary for specialized models like large language or computer vision models.
In fact, most of the interventions we curated are designed to be  generic, which can also complicate the decision-making process.
For example, if an engineer is working on a healthcare solution they may be most concerned with whether a given fairness intervention would provide adequate and relevant support for that context.

\section{Threats to validity}
\label{validity}
Our study aims to investigate the landscape of ethical AI/ML repositories within the open source community. However, there are potential threats that could impact the validity of our work.\\

\noindent\textit{External Validity.}
 One potential threat to the validity of our findings is that the landscape of fairness interventions we examined is restricted to public GitHub repositories.  \\
 
\noindent\textit{Internal Validity.}
Limiting our analysis to the top 10 search results per keyword might have overlooked valuable repositories related to ML fairness interventions. Much of the analysis in this study is based on documentation, Research Literature, and README files, which may have resulted in overlooking other artifacts that could offer deeper insights.\\

\noindent\textit{Construct Validity.}
The analysis conducted by the co-authors might overlook key insights related to the types of commonly used algorithms and metrics across the tools. Additionally, our self-constructed definition of distinguishing factors may fail to capture some significant insights.\\

\section{Conclusion}
\label{conclusion}
In this paper, we introduced a comprehensive analysis of machine learning fairness interventions, developed through an analysis of their features and functionalities. By examining the documentation and tutorials available for existing fairness interventions, we categorized our findings based on the factors a practitioner might consider when choosing or evaluating a intervention. From intervention specifications to the overarching workflows they support, our work offers valuable, practical insights into the landscape of fairness intervention support. With these perspectives, we aim to conduct interviews with fairness intervention practitioners to gain a deeper understanding of how these factors influence their selection of open source fairness interventions.

\bibliographystyle{ieeetr}
\bibliography{saner}

\end{document}